\documentclass{amsart}
\usepackage{amssymb}
\usepackage{amsfonts}
\usepackage{amsmath}
\usepackage{graphicx}%
\usepackage{lastpage}
\usepackage[utf8]{inputenc}
\usepackage[legalpaper,bookmarks=true,colorlinks=true,linkcolor=blue,citecolor=blue]{hyperref}
\usepackage{fancyhdr}
\usepackage{color}
\usepackage[mathlines]{lineno}
\usepackage{multirow}
\usepackage{caption}
\usepackage{subcaption}
\usepackage{lscape}
\usepackage{epsfig}
\usepackage{natbib}
\usepackage{float}
\usepackage{slashbox}
 \usepackage{tikz}

\theoremstyle{plain}

\newtheorem{definition}{Definition}

\newtheorem{proposition}{Proposition}

\numberwithin{equation}{section}

\begin{document}

%

\title[Joint entropies estimation : discrete case
]{Joint, Renyi-Stallis entropies and  mutual information and asymptotic limits.}

\maketitle
\author{Amadou Diadie {\sc Ba}$^{(1)}$}, Gane Samb {\sc Lo}$^{(1,2,3)}$, Cheikh Tidiane  Seck$^{(4)}$.\\

\bigskip 

\noindent \small $^{(1)}$ LERSTAD, Université Gaston Berger, Sénégal,\\
\small $^{(2)}$ Associate Researcher, LASTA, Pierre et Marie  University, Paris, FRANCE\\
\small $^{(3)}$ Assiated Professor, African University of Sciences and Technology, Abuja, NIGERIA.\\
\small $^{(4)}$ Université Alioune Diop de Bambey, Sénégal.\\

\noindent Correspondence : Amadou Diadie Ba, \email{ba.amadou-diadie@ugb.edu.sn}


 \begin{abstract}
This paper proposes a new method for estimating the 
joint probability mass function of a pair of discrete random variables. This estimator is used to construct 
 joint Shannon R\'enyi-Tsallis entropies, and the mutual information estimates of a pair of discrete random variables. 
Almost sure consistency and central limit Theorems are established. 
  Our theorical results are validated by
simulations.

\end{abstract}



\maketitle
 
 \section{Introduction}
 \subsection{Motivation} 
 \noindent 
Let $X$ and $Y$ be two discrete random variables defined on a same probability space $(\Omega, \mathcal{A}, \mathbb{P})$, with respectives values $x_1,\cdots,x_r$ and $y_1,\cdots,y_s$ (with $r> 1$ and $s>1$). \\
The \textit{information amount} of (or \textit{content} in) the outcome $(X=x_i,Y=y_j)$ is (see \cite{carter}) $$\mathcal{I}(X=x_i,Y=y_j)=\log_2 \frac{1}{p_{i,j}}$$where $ p_{i,j}=\mathbb{P}(X=x_i,Y=y_j)$.\\

\noindent The joint probability
distribution $\textbf{p}_{(X,Y)}=(p_{i,j})_{(i,j)\in I\times J}
$ of the events \par
 $(X=x_i,Y=y_j)$, coupled with the \textit{information
amount} of every event,\par  $ \mathcal{I}(X=x_i,Y=y_j)$, forms a random variable whose
expected value is the \textit{average amount of information}, or
\textit{joint entropy} (more specifically, \textit{joint Shannon entropy}), generated by this joint distribution.
\begin{definition}
Let  $X$ and $Y$ be two discrete random variables defined on a probability space $(\Omega, \mathcal{A}, \mathbb{P})$, taking respectives values in the finite countable spaces \par $X(\Omega)=\{x_1,x_2,\cdots,x_r\}$ and $Y(\Omega)=\{ y_1,\cdots,y_s\}$ (with $r> 1$ and $s>1$), and with joint probabilities mass function (p.m.f.) 
$\textbf{p}_{(X,Y)}=(p_{i,j})_{(i,j)\in I\times J}$, that is,  $$ p_{i,j}=\mathbb{P}(X=x_i,Y=y_j)\ \ \forall (i,j)\in I\times J= \{1,\cdots,r\}\times\{1,\cdots,s\}.$$

\noindent (1) The \textit{joint Shannon entropy} (JSE) of the (ordered) pair of random variables $(X,Y)$ is given by 
\begin{equation}\label{joint-shan-def}
 H(X,Y)=\sum_{(i,j)\in I\times J}p_{i,j}\log \frac{1}{p_{i,j}}=\mathbb{E}_{X,Y}\left[\log _2\frac{1}{\textbf{p}_{(X,Y)}}\right].
 \end{equation}
 \end{definition}
 \noindent Entropy is usually measured in \textit{bit}s (\textbf{b}inary \textbf{i}nformation uni\textbf{t}) (if $\log_2$), nats (if natural $\log$), or hartley( if $\log_{10}$), depending on the base of the logarithm which is used to define it.\\
 
 \noindent For ease of computations and notation convenience, we use the natural logarithm, since logarithms of varying bases are related by a constant.
\\

\noindent 
In what follows, $\textbf{p}_X=(p_{X,i})_{i\in I}$ and $\textbf{p}_Y=(p_{Y,j})_{j\in J}$
will (typically) denote the marginal distributions of the bivariate variable $(X,Y)$ whose distribution is denoted by $\textbf{p}_{(X,Y)}=(p_{i,j})_{(i,j)\in I\times J}$. Additionally, entropies will be considered as functions of p.m.f.'s, 
since they only take into account probabilities of observing specific events.\\

\noindent
 Note that 
  over all pair of random variables $(X,Y)$ that take on
at most $rs$ values with positive probability, the ones with the largest entropy are those which
are uniform on their ranges, and these random variables have entropy exactly $ \log rs$ \textit{viz}
  \begin{eqnarray*}
 H(\textbf{p}_{(X,Y)})\leq \log (rs).
 \end{eqnarray*}

 \bigskip \noindent Inspired by the study of \textit{$\alpha$-deformed algebras} and
special functions, various generalizations have been investigated. \\

\noindent Most notably, \cite{ren2} proposed a one parameter family of
entropies extending \textit{Shannon entropy}.\\

\bigskip \noindent (b) The $\alpha-$\textit{joint R\'enyi entropy} (JRE) of the pair of random variables $(X,Y)$ is defined as 
 \begin{equation}\label{joint-reyalp-def}
R_\alpha(\textbf{p}_{(X,Y)})=\frac{1}{1-\alpha} \log  \sum_{(i,j)\in I\times J}( p_{i,j})^\alpha,
 \end{equation}
\noindent with $\alpha>0,\ \ \alpha\neq 1$, which, in particular, reduces to the \textit{joint Shannon entropy} in the limit $\alpha\rightarrow 1$.\\

\bigskip \noindent (c) Also, the $\alpha-$\textit{joint Tsallis entropy} (JTE) of the pair of random variables $(X,Y)$ defined by 
\begin{eqnarray}\label{joint-tsalp-def}
T_\alpha(\textbf{p}_{(X,Y)})&=&
\frac{1}{1-\alpha}\left( \sum_{(i,j)\in I\times J}( p_{i,j})^\alpha-1\right),\ \ \alpha>0,  \ \ \alpha\neq 1
\end{eqnarray} has generated a large burst of research activities.\\

%
%

\bigskip \noindent (d) The \textit{mutual information} (MI) of the pair of random variables $(X,Y)$ defined by \begin{equation}\label{mut-def}
 I(\textbf{p}_{(X,Y)})=\sum_{(i,j)\in I\times J}p_{i,j}\log \frac{p_{i,j}}{p_{X,i}\,p_{Y,j}},
 \end{equation}
 represents the
amount of information that $Y$ reveals about $X$ (or vice versa).  
 \\ 
Here $p_{X,i}=\sum_{j=1}^{s}p_{i,j}$ and $p_{Y,j}=\sum_{i=1}^{r}p_{i,j}$
.
\vspace{2cm}
\begin{figure}

\begin{center}

   \begin{tikzpicture}
  \tikzset{venn circle/.style={draw,circle,minimum width=6cm,fill=#1,opacity=0.4,text opacity=1}}
  \node [venn circle = red] (A) at (0,0) {$H(\textbf{p}_{(X|Y)})$};
  \draw[<-] (-2,2.25) to (-3,4) 
   node[ above]{$H(\textbf{p}_X)$};
  \node [venn circle = green] (C) at (0:4cm) {$H(\textbf{p}_{(Y|X)})$};
  \draw[<-] (5.52,2.52) to (6.5,4) 
   node[ above]{$H(\textbf{p}_Y)$};
  \node[below] at (barycentric cs:A=1/2,C=1/2 ) {};   
  \node[below] at (barycentric cs:A=1,
  C=1 ) (endpoint) {$I(\textbf{p}_{(X,Y)})$};
\draw (2,-4)node[above]{$ \underbrace{\, \ \ \ \ \ \ \ \ \ \ \ \ \ \ \ \ \ \ \  \ \ \ \ \ \ \ \ \ \ \ \ \ \ \ \ \ \ \  \ \ \ \ \ \ \ \ \ \ \ \ \ \ \ \ \ \ \   \ \ \ \ \ \ \ \ \ \ \  \ \ \ \ \ \ \ \ \ \ \ \ \ \ \ \ \ \ \ }$};
\draw (2,-4.5)node[above]{\text{Union} $= H(\textbf{p}_{(X,Y)})$};
\end{tikzpicture}

\end{center}
  \caption{Diagram depicting mutual information and entropy. 
 The area contained by both circles is the joint entropy $H(\textbf{p}_{(X,Y)})$. The circle on the left (red and green) is the individual entropy $H(\textbf{p}_X)$, with the red being the conditional entropy $H(\textbf{p}_{(X|Y)})$. The circle on the right (green and red) is $H(\textbf{p}_Y)$, with the green being $H(\textbf{p}_{(Y|X)})$. The common area between $H(\textbf{p}_X)$ and $H(\textbf{p}_Y)$ at the middle is the mutual information $I(\textbf{p}_{(X,Y)})$.
}\label{rel}
 \end{figure}
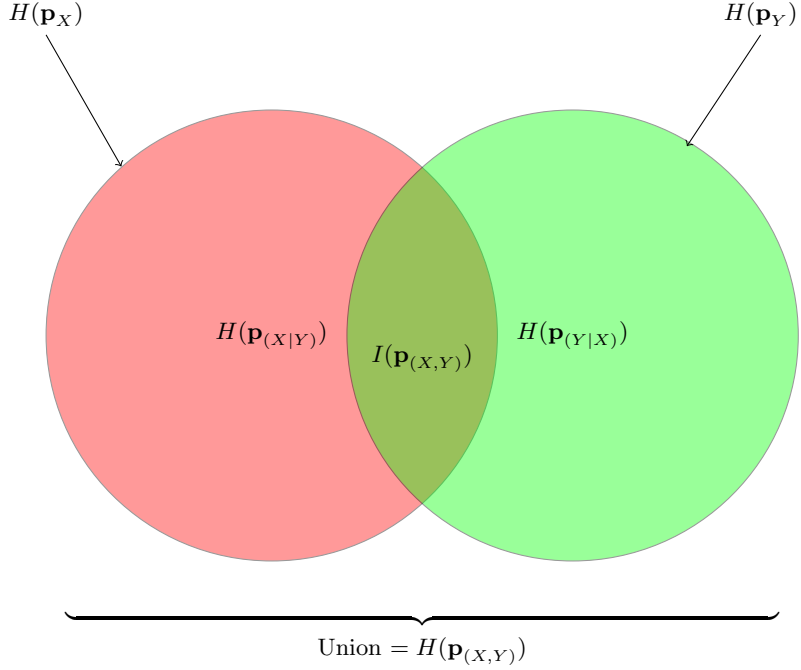

 \noindent \noindent In what follows, $\alpha>0,\ \ \alpha\neq 1$. An important relation between JRE, JTE and 
 the \textit{joint power sum} (JPS) is 
 \begin{eqnarray}\label{jre_salp}
 R_\alpha(\textbf{p}_{(X,Y)})&=&\frac{1}{1-\alpha} \log \mathcal{S}_\alpha(\textbf{p}_{(X,Y)})\\
\label{jte_salp}  \text{and}\ \  T_\alpha(\textbf{p}_{(X,Y)})&=&\frac{1}{1-\alpha} \left( \mathcal{S}_\alpha(\textbf{p}_{(X,Y)})-1\right),
 \end{eqnarray}
 where  
\begin{equation}\label{ialpha}\mathcal{S}_\alpha(\textbf{p}_{(X,Y)})=\sum_{(i,j)\in I\times J}( p_{i,j})^\alpha.
 \end{equation}

 \noindent  Mutual information is closely related to entropy 
 by
 \begin{equation}\label{chain_rule}
 I(\textbf{p}_{(X,Y)})=H(\textbf{p}_X)+H(\textbf{p}_Y)-H(\textbf{p}_{(X,Y)})
 ,
 \end{equation}where 
$$H(\textbf{p}_X)=\sum_{i\in I}( p_{X,i})^\alpha,\ \\ \ \ 
$$is 
the entropy 
 of $X$ and similarly for $Y$.\\
 This form can also be used for a Venn-diagram, as shown in \textsc{Figure} \ref{rel}
.\\

\bigskip  \noindent In this paper, our aim is to estimate directly entropies defined before by using a plug-in approach. 
\noindent \eqref{chain_rule} allows to obtain an estimator for MI by estimating $H(\textbf{p}_X)$, $H(\textbf{p}_Y)$, and $H(\textbf{p}_{(X,Y)})$ and adding them up. This corresponds to the $3H$-principle upon which number of plug-in estimators are based (see \cite{kras} for precisions on this principle). \\
 
%

\noindent In contrast, we propose in this paper  a plug-in approach that is essentially based on the estimation of the  joint probability distribution $\textbf{p}_{(X,Y)}
$ 
 from which, we can calculate the marginal distributions $\textbf{p}_X=( p_{X,i})_{(i\in I)}$, $\textbf{p}_Y=( p_{Y,j})_{(j\in J)}$ and then the quantities $H(\textbf{p}_{(X,Y)})$, $ R_\alpha(\textbf{p}_{(X,Y)})$, $
 T_\alpha(\textbf{p}_{(X,Y)})$, and $I(\textbf{p}_{(X,Y)})$. \\
 \noindent This approach is motived by the fact that studying the joint probability distribution $\textbf{p}_{(X,Y)}$ of the pair of discrete random variables $(X,Y)$ taking values, resp., in the finite sets $\mathcal{X}=\{x_i, i=1,\cdots,r\}$ and $\mathcal{Y}=\{y_j, j=1,\cdots,s\}$ is equivalent to studying the probability distribution of the $rs$ mutually exclusive possible values $(x_i,y_j)$ of $(X,Y)$. This allows us to transform the problem of estimating the joint discrete distribution of the pair $(X,Y)$ into the problem of estimating a simple distribution, say $\textbf{p}_Z$, of a single discrete random variable $Z$ suitably defined. Given an i.i.d sample of this latter random variable $Z$, we shall take,  as an estimator of the law $p_Z$, the associated empirical measure and plug it into formulas \eqref{joint-shan-def}, \eqref{joint-reyalp-def}, \eqref{joint-tsalp-def}, and \eqref{mut-def} to obtain estimates of entropies concerned.
 \\ 
 
%
%


\bigskip \noindent Before going to our entropies estimation, let highlight some important applications of them. The importance of information measures transcends information theory. Indeed, since shortly after
their inception, a wide variety of experimental sciences have found significant applications for joint Shannon entropy, Reyni and Tsallis entropies, and mutual information. For example,
\begin{itemize}
\item[$\bullet $] Finance \cite{phil};
\item[$\bullet$]Machine learning  \cite{moon2};
\item[$\bullet$] Biological sciences \cite{timm}-\cite{krishn};
\item[$\bullet$] Statistics 
\cite{liu}-\cite{lewi}-\cite{pal}-\cite{chri};
\item[$\bullet$] Sociology \cite{dave};
\item[$\bullet$] Neuroscience \cite{fred}-\cite{schne}. 

\end{itemize}

Frequently, in those applications, the need arises to estimate information measures empirically :
data are generated under an unknown probability law, and we would like to estimate these information measures from these ones.\\

\subsection{Previous work}
\noindent \textit{mutual information} estimation from samples remains an active research problem (see \cite{walt}, \cite{khan},  and \cite{sric}, to cite a few).\\
%
%
%
 
 \noindent  \cite{antos} 
defined estimator for mutual information of discrete random variables $X$ and $Y$ and  showed that,

\begin{eqnarray*}
&&\lim_{n\rightarrow +\infty}I(\widehat{ \textbf{p}}_{(X,Y)}^{(n)})\stackrel{a.s.}{=}   I(\textbf{p}_{(X,Y)})\\
\text{ and}\ \ &&\lim_{n\rightarrow +\infty}\mathbb{E}\left( I(\widehat{ \textbf{p}}_{(X,Y)}^{(n)})- I(\textbf{p}_{(X,Y)})\right)^2=0
\end{eqnarray*}
 provided that $ I(\textbf{p}_{(X,Y)})<\infty$.\\

\noindent \cite{deem}, using the 
 histogram method and under appropriate assumptions on the tail behavior of the random 
variables, showed that the \textit{mutual information} estimate 
is consistent in probability, that is, for any  $\varepsilon>0$, 
\begin{equation}\label{cvpro}
\lim_{n\rightarrow +\infty}\mathbb{P}\left(\left\vert I(\widehat{ \textbf{p}}_{(X,Y)}^{(n)})-I(\textbf{p}_{(X,Y)})\right\vert>\varepsilon \right)=0.
\end{equation}

\noindent This result will also be established by \cite{gaow} using the \textit{Kraskov–Stogbauer
–Grassberger}
(KSG) method and with some regular and smoothness conditions on resp. the Radon-Nikodym derivatives of $X$ and $Y$ and on the joint p.d.f. $ \textbf{p}_{(X,Y)}$ and with assumptions on the joint  entropy $H(\textbf{p}_{(X,Y)})$.\\

\noindent \cite{gao}, using the \textit{Local Gaussian Density Estimation} method,  proved that the \textit{mutual information} estimate is asymptotically unbiaised that is
 $$\lim_{n\rightarrow+\infty}\mathbb{E}\left(I(\widehat{ \textbf{p}}_{(X,Y)}^{(n)}) \right)= I(\textbf{p}_{(X,Y)}).$$
 
 \bigskip \noindent 

\noindent 
 By the \textit{$k-$nearest neighbors} (K-NN) method, \cite{gao} defined novel estimator for \textit{mutual information} of  mixture 
of random variables $(X,Y)$.
 They proved that the proposed estimator is asymptotically unbiaised that is 
  \begin{eqnarray*}
&& \lim_{n\rightarrow+\infty}\mathbb{E}\left(I(\widehat{ \textbf{p}}_{(X,Y)}^{(n)}) \right)= I(\textbf{p}_{(X,Y)}),
\end{eqnarray*}  provided that $k=k(n)\rightarrow+\infty$ and $(k(n)\log n)/n\rightarrow 0 $ as $n\rightarrow \infty.$\\

\noindent Furthermore, they proved that, 
if in addition $ (k(n)\log n)^2/n\rightarrow 0 $ as $n\rightarrow \infty$, then $$ \lim_{n\rightarrow+\infty}\mathbb{V}\text{ar}\left(I(\widehat{ \textbf{p}}_{(X,Y)}^{(n)}) \right)=0.$$ 

\bigskip \noindent \cite{goeb} established by Taylor approximation that, in case of independence of the two random variables $X$ and $Y$
, then 
$$ I(\textbf{p}_{(X,Y)})= \frac{1}{2\log 2}\sum_{(i,j)\in I\times J}\frac{(p_{i,j}-p_{X,i}\,p_{Y,j})^2 }{p_{X,i}\, p_{Y,j}} $$
is a second-order approximation of the \textit{mutual information}.\\
 \noindent Then they deduced that 
if 
 $ I(\textbf{p}_{(X,Y)})$ is small enough, ($<0.2$ bit) i.e. $X$ and $Y$ are independent or weakly associated random variables  and $n$ sufficiently large ($n>50$) then $I(\widehat{\textbf{p}}_{(X,Y)}^{(n)})$ approximately follows a gamma distribution with
parameters $\alpha=\frac{(r-1)(s-1)}{2}$ and $\beta=\frac{1}{n\log 2}$. \\

%
\noindent In this case the mean and variance are given as
\begin{eqnarray*}
\mathbb{E}\left(I(\widehat{ \textbf{p}}_{(X,Y)}^{(n)}) \right)&=&\frac{(r-1)(s-1)}{2n\log 2}\  \ \text{and}\ \  
\mathbb{V}\text{ar} \left(I(\widehat{ \textbf{p}}_{(X,Y)}^{(n)}) \right)=\frac{(r-1)(s-1)}{2n^2(\log 2)^2}.
\end{eqnarray*}

\bigskip \noindent \cite{xian} used the \textit{Jackknife} approach of the kernel with equalized bandwidth to estimate the \textit{S.m.i} for a pair of discrete random variables and mixed random variables (with neither purely continuous distributions nor
purely discrete distributions). \\

\bigskip
\noindent \cite{bekn} studied the \textit{mutual information} estimation for mixed pair random variables.  They developpped a kernel method to estimate the mutual information between the two random variables. The estimates enjoyed a central limit theorem under some regular conditions on the distributions.\\

%
%
%
%
%


 \subsection{ Overview of the paper} 
 The rest of the paper is organized as follows. In section \ref{maincontrib}, we define the auxiliary random variable $Z$ whose law is exactly the joint law of $(X,Y)$.
In section  \ref{estimation}, we construct plug-in estimates of joint \textit{p.m.f.}'s of $(X,Y)$ and estimates of JSE, JRE, JTE, and of MI. Section \ref{main-res}  establishes
consistency and asymptotic normality properties  of the estimates. Section \ref{testindep} is devoted to an independence test based on mutual information.
In section \ref{simulat}  we provide a simulation study to assess the performence of our estimators and we finish by a conclusion in section \ref{conclus}.

\section{Construction of the random variable $Z$ with law $\textbf{p}_{(X,Y)}$}
\label{maincontrib}

\noindent  Let $X$ and $Y$ two discrete random variables defined in the same probability space $(\Omega,\mathcal{A},\mathbb{P})$ and taking the following values $$x_1,x_2,\cdots,x_r\ \ \text{and}\ \ y_1,y_2,\cdots,y_s$$ resp. ($r>1$ and $s>1$).\\

  \noindent In addition let $Z$ a random variable defined on the same probability space $(\Omega,\mathcal{A},\mathbb{P})$ and taking the following values : $$z_1,z_2,z_3,z_4\cdots,z_{rs}.$$\\
 \noindent Denote $K=\{1,2,3,4\cdots,rs\}$.\\
 
 \noindent Simple computations give that for any $(i,j)\in I\times J$, we have $s(i-1)+j=\delta_i^j\in K $ and conversely for any $k\in K$ we have 
\begin{equation}\label{convers}
\left(1+\lfloor\frac{k-1}{s}\rfloor,k-s\lfloor \frac{k-1}{s}\rfloor \right)\in I\times J,
\end{equation} 
 where $ \lfloor x\rfloor$ denotes the largest integer less or equal to $x$.\\

\noindent 
For any possible joint values $(x_i,y_j)$ of the ordered pair $(X,Y)$, we assign the single value 
  $z_{\delta_i^j} $ of $Z$ 
 such that 
 \begin{equation}\label{pijzk}
 \mathbb{P}(X=x_i,Y=y_j)=\mathbb{P}\left(Z=z_{\delta_i^j}\right),\ \ \text{where}\ \ \delta_i^j=s(i-1)+j,
 \end{equation}and conversely, for any possible value $z_k$ of $Z$, is assigned the single pair of values $\left( x_{1+\lfloor\frac{k-1}{s}\rfloor},y_{k-s\lfloor \frac{k-1}{s}\rfloor}
\right) $ such that   
 \begin{equation}\label{zkpij}
 \mathbb{P}(Z=z_k)=\mathbb{P}\left(X=x_{1+\lfloor\frac{k-1}{s}\rfloor},Y=y_{k-s\lfloor \frac{k-1}{s}\rfloor}\right).
 \end{equation}

\noindent This means that for any $(i,j)\in I\times J$, we have 
\begin{equation}
\label{pij3}
p_{i,j}= p_{Z,s(i-1)+j}
\end{equation} where $ p_{Z,k}=\mathbb{P}(Z=z_k)$ and conversely, for any $k\in K$
\begin{equation}\label{pzk2} 
p_{Z,k}=p_{1+\lfloor\frac{k-1}{s}\rfloor,k-s \lfloor\frac{k-1}{s}\rfloor}.
\end{equation}

\bigskip 
\noindent \textsc{Table} \ref{tabjpd} illustrates the correspondance between $p_{i,j}$ and $p_{Z,k}$, for ($i,j,k)\in I\times J\times K$.

\bigskip 
\noindent From there, the marginals \textit{p.m.f.'}s  $p_{X,i}$ 
 are expressed from \textit{p.m.f.}'s of the random variable $Z$ by 
 \begin{eqnarray}
\label{pij2} p_{X,i}&=&\sum_{j=1}^sp_{Z,\delta_i^j},\ \
 \ p_{Y,j}=\sum_{i=1}^rp_{Z,\delta_i^j}
.
 \end{eqnarray}

\begin{center}
\vspace{9ex}
\begin{table}
\centering
\begin{tabular}{lllll}
$p_{1,1}=p_{Z,1}$ & $\cdots$ & $p_{1,j}=p_{Z,j}$ &$ \cdots$ &$ p_{1,s}=p_{Z,s}$ \\ 
$p_{2,1}=p_{Z,s+1}$ &$\cdots$ &$ p_{2,j}=p_{Z,s+j}$ &$ \cdots $& $ p_{2,s}=p_{Z,2s}$ \\ 
$\vdots $&$ \vdots$ &$ \vdots $&$ \vdots $&$ \vdots$ \\ 
$p_{i,1}=p_{Z,s(i-1)+1} $&$ \cdots$ & $p_{i,j}=p_{Z,\delta_i^j}$ &$ \cdots $&$ p_{i,s}=p_{Z,si}$ \\ 
$\vdots$ &$ \vdots $&$ \vdots $&$ \vdots $&$ \vdots$ \\ 
$p_{r,1}=p_{Z,s(r-1)+1}$ &$ \cdots $&$ p_{r,j}=p_{Z,s(r-1)+j}$&$\cdots$ & $p_{r,s}=p_{Z,rs}$
\end{tabular} 
$$\text{conversely}$$
$$\begin{array}{cccccc}
p_{Z,1}=p_{1,1} & p_{Z,2}=p_{1,2} & \cdots & p_{Z,k}=p_{1+\lfloor\frac{k-1}{s}\rfloor,k-s\lfloor \frac{k-1}{s}\rfloor
} 
& \cdots & p_{Z,rs}=p_{r,s} \end{array} $$
\vspace{3ex} 
\caption{Illustration of the correspondance between $\textbf{p}_{(X,Y)}$ and $\textbf{p}_Z$.
}\label{tabjpd}
\end{table}
\end{center}
%
%
%

\noindent Finally, JSE, JRE, JTE and MI are expressed simply in terms of $\textbf{p}_Z=(p_{Z,k})_{k\in K}$ through (\ref{pij3}), that is 
\begin{eqnarray*}
&& H(\textbf{p}_{(X,Y)})= -\sum_{k\in K}p_{Z,k}\log p_{Z,k},\ \ 
  R_\alpha\left(\textbf{p}_{(X,Y)}\right)=\frac{1}{1-\alpha}\log \left(\sum_{k\in K}(p_{Z,k})^\alpha\right),\\ \ \
&&  T_\alpha\left(\textbf{p}_{(X,Y)}\right)=\frac{1}{1-\alpha}\left(\sum_{k\in K}(p_{Z,k})^\alpha-1\right),\\
&& \text{and}\ \ I(\textbf{p}_{(X,Y)})=\sum_{(i,j)\in I\times J}p_{Z,\delta_i^j}\log \frac{p_{Z,\delta_i^j}}{p_{X,i}\,p_{Y,j}},
\end{eqnarray*}where $\alpha>0,\ \ \alpha\neq 1$.

\bigskip 

\noindent We may give now the following remark : \\
 
   \noindent  For most of univariate or multivariate entropies, we may have computation problems. So without loss of generality, suppose
\begin{equation}
\ \ \ p_{Z,k}>0,\ \ \ \forall k\in K
 \ \ \ \ \ (\textbf{BD}).
\label{BD}
\end{equation}

\bigskip \noindent If Assumption (\ref{BD}) holds, we do not have to worry about summation problems. This explain why Assumption \eqref{BD} is systematically used in a great number of works in that topics, for example, in \cite{hall}, \cite{singh}, \cite{kris}, and recently \cite{ba5}, to cite a few.

\section{Estimation} \label{estimation}


\noindent In this section, we construct estimate of \textit{p.m.f.} $\textbf{p}_{Z,k}$ from i.i.d. random variables according to $\textbf{p}_Z$, and we give some inescapable results needed in the sequel, and finally construct the plug-in estimates of the entropies cited above. 
 \\
 
 \noindent Let $Z_1,\cdots,Z_n$ be $n$ i.i.d. random variables from $Z$ and according to $\textbf{p}_Z$. \\

\noindent Here, it is worth noting that, in the sequel, $K=\{1,2,\cdots,rs\}$, with $r$ and $s$ integers strictly greater than $1$. This means that 
    $rs$ can not be a prime number so that \eqref{pzk2} holds. \\

\noindent  For a given $k\in K$, define the easiest and most objective estimator of $p_{Z,k}$, based on the i.i.d sample $ Z_1,\cdots,Z_n,$ by 
 \begin{eqnarray}\label{pn}
 \widehat{p}_{Z,k}^{(n)}
 &=&\frac{1}{n}\sum_{\ell=1}^n1_{z_k}(Z_\ell)
\end{eqnarray}where 
 $1_{z_k}(Z_\ell)=\begin{cases}
 1\ \ \text{if}\ \ Z_\ell=z_k\\
 0\ \ \text{otherwise}
 \end{cases} $ for a fixed  $k\in K$.
\\

\noindent This means that, for a given $(i,j)\in I\times J$, an estimate of $p_{i,j}$ based on the i.i.d sample $ Z_1,\cdots,Z_n,$ according to $\textbf{p}_Z$ is given by  
\begin{equation}\label{pxyni}
 \widehat{p}_{i,j}^{(n)}= \widehat{p}_{Z,\delta_{i}^j}^{(n)}=\frac{1}{n}\sum_{\ell=1}^n1_{z_{\delta_i^j}}(Z_\ell)
 .
\end{equation}
\noindent 
where $1_{z_{\delta_i^j}}(Z_\ell)=\begin{cases}
1\ \ \text{if}\ \ Z_\ell=z_{\delta_i^j}\\
 0\ \ \text{otherwise}
\end{cases}$ for fixed $(i,j)\in I\times J$.\\

\bigskip \noindent From \eqref{pij2}, estimate of each of the marginals pdf's  $p_{X,i}$ and $ p_{Y,j}$ are 
\begin{eqnarray}\label{pxni}
\widehat{p}_{X,i}^{(n)}&=& \sum_{j=1}^s\widehat{p}_{Z,\delta_{i}^j}^{(n)}=\frac{1}{n}\sum_{\ell=1}^n1_{A_i}(Z_\ell)\\
\text{and} \ \ \ 
\widehat{p}_{Y,j}^{(n)}&=& \sum_{i=1}^r\widehat{p}_{Z,\delta_{i}^j}^{(n)}=\frac{1}{n}\sum_{\ell=1}^n1_{B_j}(Z_\ell),\label{pynj}
\end{eqnarray}
with 
\begin{eqnarray*}
A_i&=&\{z_{s(i-1)+1},z_{s(i-1)+2},\cdots,z_{si}\}=\bigcup_{j=1}^s \{z_{\delta_i^j}\}\\ \text{and}\ \ B_j&=&\{z_{j},z_{s+j},z_{2s+j},\cdots,z_{s(r-1)+j}\}=\bigcup_{i=1}^r\{ z_{\delta_i^j}\}.
\end{eqnarray*}

\bigskip \noindent In the following, we use equally $p_{Z,k}$ or $p_{i,j}$  since they are equal in consideration of \eqref{pij3} and \eqref{pzk2} and we denote
\begin{eqnarray}\label{dist_pij}
\widehat{ \textbf{p}}_{(X,Y)}^{(n)}=(\widehat{p}_{i,j}^{(n)})_{(i,j)\in I\times J},\ 
 \ \  \widehat{\textbf{p}}_{X}^{(n)}=(\widehat{p}_{X,i}^{(n)})_{i\in I}\ \ \ \text{and}\ \ \widehat{\textbf{p}}_{Y}^{(n)}=(\widehat{p}_{Y,j}^{(n)})_{j\in J}.
\end{eqnarray}

\noindent Before going further, let give some results concerning the empirical estimator \eqref{pn}. 
\\

\noindent For a given $k\in K$, this empirical estimator $\widehat{p}_{Z,k}^{(n)}$ is strongly consistent and asymptotically normal. Precisely, for a fixed $k\in K$, when $n$ tends to infinity,
 \begin{eqnarray}
\label{pzn}
&&\widehat{p}_{Z,k}^{(n)}- p_{Z,k} \stackrel{a.s.}{\longrightarrow} 0,\\
&&\label{pnj}  \sqrt{n}(\widehat{p}_{Z,k}^{(n)}- p_{Z,k})  \stackrel{\mathcal{D}}{\rightsquigarrow}G_{p_{Z,k}}.
\end{eqnarray}  where $G_{p_{Z,k}}\stackrel{d }{\sim}\mathcal{N}(0, p_{Z,k} (1-p_{Z,k} ))$.\\ 

\noindent These asymptotic properties derive from the law of large
numbers and central limit theorem. \\
 
\bigskip \noindent Here and in the following, $\stackrel{a.s.}{ \longrightarrow}$ means the \textit{almost sure convergence},  $\stackrel{\mathcal{D}}{ \rightsquigarrow}
$, the \textit{convergence in distribution}, and $\stackrel{d }{\sim}$, means \textit{equality in distribution}. \\

\noindent Recall that,
since for a fixed $k\in K,$ $n\widehat{p}_{Z,k}^{(n)}$ has a binomial distribution with parameters $n$  and success probability $p_{Z,k}$, we have 
 \begin{equation*}
 \mathbb{E}\left[ \widehat{p}_{Z,k}^{(n)}\right]=p_{Z,k} \ \ \text{and}\ \ \mathbb{V}\text{ar}(\widehat{p}_{Z,k}^{(n)})=\frac{p_{Z,k} (1-p_{Z,k} )}{n}.
\end{equation*}

\noindent Denote
\begin{eqnarray*}
\rho_n(p_{Z,k})=\sqrt{n/p_{Z,k}}\Delta_{p_{Z,k}}^{(n)}\ \ \text{and}\ \ a_{Z,n}=\sup_{k\in K}\left\vert  \Delta_{p_{Z,k}}^{(n)}
\right\vert,
\end{eqnarray*}where $\Delta_{p_{Z,k}}^{(n)}=\widehat{p}_{Z,k}^{(n)}- p_{Z,k}.$\\

\noindent By the asymptotic Gaussian limit of the multinomial law (see for example \cite{ips-wcia-ang}, Chapter 1, Section 4), we have
\begin{eqnarray}
\label{ropnj}&& \biggr( \rho_n(p_{Z,k}), \ k\in K\biggr)
\stackrel{\mathcal{D}}{\rightsquigarrow }G(\textbf{p}_Z),\ \ \ \ \text{as}\ \ n\rightarrow +\infty,
\end{eqnarray}where $G(\textbf{p}_Z)= (G_{p_{Z,k}},k\in K)^t\stackrel{d }{\sim}\mathcal{N}(0,\Sigma_{\textbf{p}_Z}),$ and $\Sigma_{\textbf{p}_Z}$ is the covariance matrix which elements are :
\begin{eqnarray}\label{vars}
&&\sigma_{(k,k')}=(1-p_{Z,k} )1_{(k=k')}-\sqrt{ p_{Z,k}p_{Z,k'} } 1_{(k\neq k')}, \ \ (k,k') \in K^2.
\end{eqnarray}

 \bigskip \noindent 
By denoting $
a_{X,n}=\sup_{i\in I}|\widehat{p}_{X,i}^{(n)}-p_{X,i}|\ \ \text{and}\ \ a_{Y,n}=\sup_{j\in J}|\widehat{p}_{Y,j}^{(n)}-p_{Y,j}|
$ then, we have  
\begin{equation}\label{maxaxy}
\max(a_{X,n},a_{Y,n})\stackrel{a.s.}{\longrightarrow}0 \ \ \text{as}\ \ n\rightarrow +\infty. 
\end{equation}

 \bigskip \noindent 
As a consequence, JSE,  JRE, 
 and JTE 
  are estimated from the sample $Z_1,\cdots,Z_n$ by their plug-in counterparts,
meaning that we simply insert the consistent \textit{p.m.f.} estimator $\widehat{p}_{Z,k}^{(n)}$ computed from \eqref{pn} in place of 
 JSE,  JRE, and JTE expresions \textit{viz} :
\begin{eqnarray}
H(\widehat{\textbf{p}}_{(X,Y)}^{(n)} )&=&-\sum_{(i,j)\in I\times J}\widehat{p}_{Z,\delta_i^j}^{(n)}\log \widehat{p}_{Z,\delta_i^j}^{(n)},
\ \ \\
 R_\alpha\left(\widehat{\textbf{p}}_{(X,Y)}^{(n)}
\right)&=&\frac{1}{1-\alpha}\log \left(\sum_{(i,j)\in I\times J}( \widehat{p}_{Z,\delta_i^j}^{(n)})^\alpha\right),\\
T_\alpha\left(\widehat{\textbf{p}}_{(X,Y)}^{(n)}
\right)
& =&\frac{1}{1-\alpha}\left(\sum_{(i,j)\in I\times J}( \widehat{p}_{Z,\delta_i^j}^{(n)})^\alpha-1\right),
\\
\label{smiestim}\text{and}\ \ \ \ I\left(\widehat{\textbf{p}}_{(X,Y)}^{(n)}
\right)&=&\sum_{(i,j)\in I\times J}\widehat{p}_{Z,\delta_i^j}^{(n)}\log \frac{\widehat{p}_{Z,\delta_i^j}^{(n)}
}{\widehat{p}_{X,i}^{(n)}\,\widehat{p}_{Y,j}^{(n)}
}.
\end{eqnarray}
where $\alpha>0,\ \ \alpha\neq 1$ and $\widehat{p}_{Z,\delta_i^j}^{(n)}$, and $\widehat{p}_{X,i}^{(n)},$ are given resp. by \eqref{pxyni}, and \eqref{pxni}.\\

\noindent In addition, define the JPS estimate 
\begin{eqnarray}
\mathcal{S}_\alpha\left(\widehat{\textbf{p}}_{(X,Y)}^{(n)}
\right)=
\sum_{(i,j)\in I\times J}\left( \widehat{p}_{Z,\delta_i^j}^{(n)}\right)^\alpha.
\end{eqnarray}
%
%

\bigskip \noindent In the following, we present asymptotic limits of these empirical estimators.

\section{Statements of the main results}\label{main-res}

\bigskip \noindent In this section, we state and prove almost sure consistency and central limit theorem  for the estimators defined above. \\

 \subsection{ Asymptotic limits of joint Shannon entropy estimate
.}
$\,$\\

\noindent Denote  
\begin{eqnarray}
&&\label{asz} A_H(\textbf{p}_{(X,Y)})=\sum_{k\in K}\left\vert 1+ \log  ( p_{Z,k} )\right \vert\\
\label{sigz}\ \ \ 
&&\sigma_H^2(\textbf{p}_{(X,Y)})=\sum_{k\in K} p_{Z,k}(1- p_{Z,k} )(1+ \log  ( p_{Z,k} ))^2\\
&&\notag \ \ \ \ \ \ \ \  \ - \ \ 2  \ \sum_{(k,k')\in K^2,k\neq k'}(p_{Z,k}p_{Z,k'})^{3/2}(1+\log (p_{Z,k}))(1+\log (p_{Z,k'})).\end{eqnarray}
 
\begin{proposition}
 \label{jent}Let $\textbf{p}_{(X,Y)}$ a probability distribution and $\widehat{\textbf{p}}_{(X,Y)}^{(n)}$ be generated by i.i.d samples $Z_1,Z_2,\cdots,Z_n$ according to $\textbf{p}_{(X,Y)}$ and given by \eqref{dist_pij}, assumption \eqref{BD} be satisfied.
 Then  the following asymptotic results hold
\begin{eqnarray}\label{epas}
&&\limsup_{n\rightarrow +\infty}\frac{\left\vert H(\widehat{\textbf{p}}_{(X,Y)}^{(n)})-H(\textbf{p}_{(X,Y)}) \right\vert}{ a_{Z,n} }\leq A_H(\textbf{p}_{(X,Y)}),\ \ \text{a.s.}\\
&&\sqrt{n}\left(H(\widehat{\textbf{p}}_{(X,Y)}^{(n)})-H(\textbf{p}_{(X,Y)}) \right)\stackrel{\mathcal{D}}{ \rightsquigarrow} \mathcal{N}(0,\sigma_H^2(\textbf{p}_{(X,Y)})),\ \ \text{as}\ \ n\rightarrow +\infty.\label{epnor}
\end{eqnarray}
\end{proposition}

 \begin{proof}Define the function $\psi:\,(0,+\infty)\rightarrow \mathbb{R}$ by $\psi(x)=x\log x$. \\
 \noindent Let $(i,j)\in I\times J$, and set  $k=\delta_i^j\in K$. We have
 \begin{eqnarray}\label{mvtpz}
\notag \psi( \widehat{p}_{Z,k}^{(n)})&=&\psi(p_{Z,k}+\Delta_{p_{Z,k}}^{(n)})\\
&=&\psi(p_{Z,k})+\Delta_{p_{Z,k}}^{(n)}\psi'( p_{Z,k}+\theta_{1,k}^{(n)}\Delta_{p_{Z,k}}^{(n)}),
\end{eqnarray} by the mean values theorem and where $\theta_{1,k}^{(n)}$ is some number lying in $(0,1)$.\\
\noindent Applying again the main value Theorem to the derivative function $\psi'$ of $\psi$, we obtain 
\begin{eqnarray*}
\psi'( p_{Z,k}+\theta_{1,k}^{(n)}\Delta_{p_{Z,k}}^{(n)})&=&\psi'( p_{Z,k})+\theta_{1,k}^{(n)}\Delta_{p_{Z,k}}^{(n)}\psi"(p_{Z,k}+\theta_{2,k}^{(n)}\Delta_{p_{Z,k}}^{(n)})
\end{eqnarray*} where $ \theta_{2,k}^{(n)}\in(0,1)$. Replacing in \eqref{mvtpz}, it yields 
\begin{eqnarray*}
\psi( \widehat{p}_{Z,k}^{(n)})&=&\psi(p_{Z,k})+\Delta_{p_{Z,k}}^{(n)}\psi'( p_{Z,k})+\theta_{1,k}^{(n)}(\Delta_{p_{Z,k}}^{(n)})^2\psi"(p_{Z,k}+\theta_{2,k}^{(n)}\Delta_{p_{Z,k}}^{(n)}).
\end{eqnarray*} Now summing over $(i,j)\in I\times J$, it follows that 
\begin{eqnarray}\notag
H(\widehat{\textbf{p}}_{(X,Y)}^{(n)} )-H(\textbf{p}_{(X,Y)})&=&-\sum_{(i,j)\in I\times J}\Delta_{p_{Z,\delta_i^j}}^{(n)}\psi'( p_{Z,\delta_i^j})\\
&&-\sum_{(i,j)\in I\times J}\theta_{1,\delta_i^j}^{(n)}(\Delta_{p_{Z,\delta_i^j}}^{(n)})^2\psi"(p_{Z,\delta_i^j}+\theta_{2,\delta_i^j}^{(n)}\Delta_{p_{Z,k}}^{(n)})\label{aspz}
\end{eqnarray}
so that \begin{eqnarray*}
|  H(\widehat{\textbf{p}}_{(X,Y)}^{(n)} )-H(\textbf{p}_{(X,Y)})|\leq a_{n,Z}\sum_{(i,j)\in I\times J}|\psi'( p_{Z,\delta_i^j})|+(a_{n,Z})^2\sum_{(i,j)\in I\times J}|\psi"(p_{Z,\delta_i^j}+\theta_{2,\delta_i^j}^{(n)}\Delta_{p_{Z,k}}^{(n)})|.
\end{eqnarray*}
\noindent Hence \begin{eqnarray*}
\limsup_{n\rightarrow+\infty}\frac{ |  H(\widehat{\textbf{p}}_{(X,Y)}^{(n)} )-H(\textbf{p}_{(X,Y)})|}{a_{n,Z}}\leq 
\sum_{k\in K}\left\vert 1+ \log  ( p_{Z,k} )\right \vert,\ \ \text{a.s.},
\end{eqnarray*}since, as $n\rightarrow+\infty$, 
$$ \psi"(p_{Z,\delta_i^j}+\theta_{2,\delta_i^j}^{(n)}\Delta_{p_{Z,k}}^{(n)}) \rightarrow \psi"(p_{Z,\delta_i^j})<\infty.$$

\noindent Which proves the claim \eqref{epas}.\\

\noindent Going back to \eqref{aspz}, we have

\begin{eqnarray}\notag
\sqrt{n}( H(\widehat{\textbf{p}}_{(X,Y)}^{(n)} )-H(\textbf{p}_{(X,Y)}))
\label{anpz}&=&-\sum_{k\in K}\sqrt{p_{Z,k}}\rho_n(p_{Z,k})\psi'( p_{Z,k})+\sqrt{n}R_{1,n}.
\end{eqnarray}where $$R_{1,n}=-\sum_{k\in K}\theta_{1,k}^{(n)}(\Delta_{p_{Z,k}}^{(n)})^2\psi"(p_{Z,k}+\theta_{2,k}^{(n)}\Delta_{p_{Z,k}}^{(n)}).$$

\noindent The asymptotic Gaussian limit of the multinomial law \eqref{ropnj}, garantees that

\begin{eqnarray*}\sum_{k\in K}\sqrt{p_{Z,k}}\rho_n(p_{Z,k})\psi'( p_{Z,k})\stackrel{\mathcal{D} }{\rightsquigarrow}%
\mathcal{N}(0,\sigma^2(\textbf{p}_Z))
,\ \ \text{as}\ \ n\rightarrow+\infty,
\end{eqnarray*}
where the asymptotic variance, $\sigma^2(\textbf{p}_Z)$, equals to 

\begin{eqnarray*}
&&\mathbb{V}\text{ar}\left(\sum_{k\in K}\sqrt{p_{Z,k}}\psi'( p_{Z,k})G_{p_{Z,k}}\right)\\
&&=\sum_{k\in K}\mathbb{V}\text{ar}\left(\sqrt{p_{Z,k}}\psi'( p_{Z,k})G_{p_{Z,k}}\right)\\
&&+2\sum_{(k,k')\in K^2,k\neq k'}\mathbb{C}\text{ov}\left( \sqrt{p_{Z,k}}\psi'(p_{Z,k})G_{p_{Z,k}},\sqrt{p_{Z,k'}}\psi'( p_{Z,k'})G_{p_{Z,k'}} \right)\\
&=&\sum_{k\in K}p_{Z,k}(\psi'( p_{Z,k}))^2\mathbb{V}\text{ar}\left(G_{p_{Z,k}}\right)\\
&&+2\sum_{(k,k')\in K^2,k\neq k'}\sqrt{p_{Z,k}} \sqrt{p_{Z,k'}} \psi'( p_{Z,k})\psi'( p_{Z,k'})\mathbb{C}\text{ov}\left(G_{p_{Z,k}},G_{p_{Z,k'}} \right)\\
&=&\sum_{k\in K} p_{Z,k}(1- p_{Z,k} )(1+ \log  ( p_{Z,k} ))^2\\
&&\ \ \ \ \ \ \ \  \ - \ \ 2  \ \sum_{(k,k')\in K^2,k\neq k'}(p_{Z,k}p_{Z,k'})^{3/2}(1+\log (p_{Z,k}))(1+\log (p_{Z,k'})).
\end{eqnarray*}

\noindent It remains to prove that $\sqrt{n}R_{1,n}$ converges in probability to $0$ as $n\rightarrow+\infty$.
\\
 \noindent We have 
 \begin{eqnarray*}
 |\sqrt{n}R_{1,n}|\leq \sqrt{n}(a_{Z,n})^2\sum_{k\in K}\left\vert \psi"(p_{Z,k}+\theta_{2,k}^{(n)}\Delta_{p_{Z,k}}^{(n)})\right\vert.
 \end{eqnarray*}
 \noindent By the Bienaymé-Tchebychev inequality, we have, for any fixed $\epsilon >0$ and for any $k\in K$
\begin{eqnarray*}
\mathbb{P}(\sqrt{n}( \widehat{p}_{Z,k}^{(n)} - p_{Z,k} )^2\geq \epsilon)=\mathbb{P}\left(|  \widehat{p}_{Z,k}^{(n)} - p_{Z,k} |\geq  \frac{\sqrt{\epsilon} }{n^{1/4}}\right)\leq \frac{p_{Z,k} (1- p_{Z,k} )}{\epsilon n^{1/2}}.
\end{eqnarray*} 
\noindent Therefore $\sqrt{n}(a_{Z,n})^{2}=o_{\mathbb{P}}(1)$ which entails that $\sqrt{n}R_{1,n}=0_{\mathbb{P}}(1)$ since, as $n$ tends to $+\infty$, we have  $$\sum_{k\in K}\left\vert \psi"(p_{Z,k}+\theta_{2,k}^{(n)}\Delta_{p_{Z,k}}^{(n)})\right\vert\rightarrow \sum_{k\in K}\left\vert \psi"(p_{Z,k})\right\vert <\infty.$$

\noindent All this proves the claim \eqref{epnor} and ends the proof of the Proposition \ref{jent}
\end{proof}

\subsection{ Asymptotic limit of joint Renyi and Tsallis entropies estimates}$\,$\\

\bigskip \noindent 
The following proposition 
concerns the asymptotic limits of JPS estimate $\mathcal{S}_\alpha(\widehat{\textbf{p}}_{(X,Y)}^{(n)} )$ given by
\begin{equation}
\label{ialpha_estim}\mathcal{S}_\alpha(\widehat{\textbf{p}}_{(X,Y)}^{(n)} )=\sum_{(i,j)\in I\times J}\widehat{p}_{Z,\delta_i^j}^{(n)},\ \ \ \alpha>0,\ \ \alpha\neq 1.
\end{equation}
The proof is the same as that of Proposition \ref{jent}, just replace the function $\psi$ by the function $ \varphi: x\mapsto x^\alpha.$ Hence omitted.\\

\bigskip \noindent For $\alpha>0,\ \ \alpha\neq 1$, denote 
\begin{eqnarray*}
A_{\mathcal{S}_\alpha}(\textbf{p}_{(X,Y)})&=& \alpha\sum_{k\in K}\left(p_{Z,k}\right)^{\alpha-1},\\
  \sigma_{\mathcal{S}_\alpha}^2(\textbf{p}_{(X,Y)})
& =&\alpha^2\biggr[ \sum_{k\in K}( p_{Z,k})^{2\alpha-1}(1-p_{Z,k})- 2 \sum_{(k,k')\in K^2,k\neq k'}\left( p_{Z,k} p_{Z,k'}\right)^{\alpha+1/2}\biggr]
.
\end{eqnarray*} 
\begin{proposition}
 \label{corSalp}Under the conditions as in Proposition \ref{jent}, the asymptotic results hold
\begin{eqnarray}\label{ialphps}
&& \limsup_{n\rightarrow+\infty}\frac{\left\vert \mathcal{S}_\alpha(\widehat{\textbf{p}}_{(X,Y)}^{(n)} )-\mathcal{S}_\alpha(\textbf{p}_{(X,Y)})\right\vert}{ a_{Z,n} }\leq A_{\mathcal{S}_\alpha}(\textbf{p}_{(X,Y)}),\ \ \text{a.s.}\\
&&\sqrt{n}\left(\mathcal{S}_\alpha(\widehat{\textbf{p}}_{(X,Y)}^{(n)} )-\mathcal{S}_\alpha(\textbf{p}_{(X,Y)})\right)\stackrel{\mathcal{D} }{\rightsquigarrow}\mathcal{N}(0,\sigma_{\mathcal{S}_\alpha}^2(\textbf{p}_{(X,Y)})),\ \ \text{as}\ \ n\rightarrow+\infty.\label{ialphno}
\end{eqnarray}

\end{proposition}

\bigskip \noindent Turning now to our second result, note that the relation \eqref{jre_salp} suggests that similar results of
Proposition \ref{corSalp} could be also extended to the JRE.\\ 
%

\noindent For $\alpha>0,\ \ \alpha\neq 1$, denote 
\begin{eqnarray*}
A_{R,\alpha}(\textbf{p}_{(X,Y)})&=&\frac{A_{\mathcal{S}_\alpha}(\textbf{p}_{(X,Y)})}{\left\vert1-\alpha\right\vert \mathcal{S}_\alpha(\textbf{p}_{(X,Y)})}\\
\text{and}\ \  \ 
\sigma _{R,\alpha}^{2}(\textbf{p}_{(X,Y)})&=&\left( \frac{1
 }{(1-\alpha)\mathcal{S}_\alpha(\textbf{p}_{(X,Y)})}\right)^2\sigma_{\mathcal{S}_\alpha}^2(\textbf{p}_{(X,Y)}).
\end{eqnarray*}

\begin{proposition}

\label{thejRen} 
Under the same assumptions as in Proposition \ref{corSalp},  the following asymptotic results hold
\begin{eqnarray}\label{jRenas}
&& \limsup_{n\rightarrow +\infty}\frac{|R_\alpha( \widehat{\textbf{p}}_{(X,Y)}^{(n)})-R_\alpha(\textbf{p}_{(X,Y)})|}{ a_{Z,n} }\leq A_{R,\alpha}(\textbf{p}_{(X,Y)}),\ \ \text{a.s}.\\
&&\label{jRenan}
\sqrt{n}\left( R_\alpha( \widehat{\textbf{p}}_{(X,Y)}^{(n)})-R_\alpha(\textbf{p}_{(X,Y)})\right)\stackrel{\mathcal{D}}{ \rightsquigarrow} 
\mathcal{N}\left( 0,\sigma _{R,\alpha}^{2}(\textbf{p}_{(X,Y)})\right)\text{ as } n\rightarrow + \infty.
\end{eqnarray}

\end{proposition}\begin{proof}
 For $\alpha\in (0,1)\cup (1,+\infty),$ 
 we have
\begin{equation}\label{eralp}
R_\alpha( \widehat{\textbf{p}}_{(X,Y)}^{(n)})-R_\alpha(\textbf{p}_{(X,Y)})=%
\frac{1}{1-\alpha }\left(  \log  \mathcal{S}_\alpha( \widehat{\textbf{p}}_{(X,Y)}^{(n)} )- \log \mathcal{S}_\alpha(\textbf{p}_{(X,Y)})\right).
\end{equation}%
Using a Taylor
expansion of $ \log  (1+y)$ it follows that almost surely, 
\begin{eqnarray*}
 \log  \mathcal{S}_\alpha( \widehat{\textbf{p}}_{(X,Y)}^{(n)} )- \log \mathcal{S}_\alpha(\textbf{p}_{(X,Y)}) &=& \log  \left( 1+%
\frac{\mathcal{S}_\alpha( \widehat{\textbf{p}}_{(X,Y)}^{(n)} ))-\mathcal{S}_{\alpha}(\textbf{p})}{\mathcal{S}_\alpha(\textbf{p}_{(X,Y)})}%
\right) \\
&=&\frac{\mathcal{S}_\alpha( \widehat{\textbf{p}}_{(X,Y)}^{(n)} )-\mathcal{S}_\alpha(\textbf{p}_{(X,Y)})}{\mathcal{S}_\alpha(\textbf{p}_{(X,Y)})}+O_{\text{a.s%
}}(a_{Z,n}^{2}).
\end{eqnarray*}
\noindent Finally this, combined with \eqref{ialphps} of Proposition \ref{corSalp}, proves the claim \eqref{jRenas}.\\

\bigskip \noindent Let prove the claim \eqref{jRenan}.  \\

\noindent Using the same technics as in the proof of Proposition \ref{jent}, we obtain 
\begin{eqnarray}
\sqrt{n}\left( \mathcal{S}_\alpha( \widehat{\textbf{p}}_{(X,Y)}^{(n)} ) -\mathcal{S}_\alpha(\textbf{p}_{(X,Y)})\right) &=&\sqrt{n} \sum_{k\in K}\Delta_{p_{Z,k}}^{(n)}\varphi'( p_{Z,k})+o_{\mathbb{P}}(1)
\label{assalp}
\end{eqnarray} where $\varphi(x)=x^\alpha$. So that dividing each member by 
$\sqrt{n} \mathcal{S}_\alpha(\textbf{p}_{(X,Y)})$, we get\begin{equation*}
\frac{ \mathcal{S}_\alpha( \widehat{\textbf{p}}_{(X,Y)}^{(n)} ))}{\mathcal{S}_\alpha(\textbf{p}_{(X,Y)})}=1+\frac{\sum_{k\in K}\Delta_{p_{Z,k}}^{(n)}\varphi'( p_{Z,k})}{\mathcal{S}_\alpha(\textbf{p}_{(X,Y)})}+o_{\mathbb{P}}(1).
\end{equation*}

\noindent Now by Taylor expansion of $ \log  (1+y)$, it follows that, almost surely, 
\begin{eqnarray*}
 \log  \mathcal{S}_\alpha( \widehat{\textbf{p}}_{(X,Y)}^{(n)} ))- \log \mathcal{S}_\alpha(\textbf{p}_{(X,Y)}) &=& \log  \left(1+\frac{\sum_{k\in K}\Delta_{p_{Z,k}}^{(n)}\varphi'( p_{Z,k})}{\mathcal{S}_\alpha(\textbf{p}_{(X,Y)})}\right) \\
&=&\frac{\sum_{k\in K}\Delta_{p_{Z,k}}^{(n)}\varphi'( p_{Z,k})}{\mathcal{S}_\alpha(\textbf{p}_{(X,Y)})}+O_{%
\mathbb{P}}\left( \frac{1}{n}\right)
\end{eqnarray*}

\noindent thus, from \eqref{eralp}, we obtain
\begin{eqnarray*}
\sqrt{n}\left(R_\alpha( \widehat{\textbf{p}}_{(X,Y)}^{(n)})-R_\alpha(\textbf{p}_{(X,Y)})\right)&=&\frac{1}{(1-\alpha)\mathcal{S}_\alpha(\textbf{p}_{(X,Y)}) }\sum_{k\in K}\sqrt{n}\Delta_{p_{Z,k}}^{(n)}\varphi'( p_{Z,k})+o_{\mathbb{P}}(1),\\ 
\end{eqnarray*}
but using \eqref{ialphno}, we have that 
\begin{eqnarray*}
\sum_{k\in K}\sqrt{n}\Delta_{p_{Z,k}}^{(n)}\varphi'( p_{Z,k})\stackrel{\mathcal{D} }{\rightsquigarrow}\mathcal{N}(0,\sigma_{\mathcal{S}_\alpha}^2(\textbf{p}_{(X,Y)})),\ \ \text{as}\ \ n\rightarrow+\infty.
\end{eqnarray*}
Finally 
\begin{eqnarray*}
\sqrt{n}\left(R_\alpha( \widehat{\textbf{p}}_{(X,Y)}^{(n)})-R_\alpha(\textbf{p}_{(X,Y)})\right)\stackrel{\mathcal{D} }{\rightsquigarrow}\mathcal{N}(0,\sigma_{R,\alpha}^2(\textbf{p}_{(X,Y)})),\ \ \text{as}\ \ n\rightarrow+\infty,
\end{eqnarray*}with
\begin{eqnarray*}
 \sigma _{R,\alpha}^{2}(\textbf{p}_{(X,Y)})&=&\left(\frac{1}{(1-\alpha)\mathcal{S}_\alpha(\textbf{p}_{(X,Y)})}\right)^2\sigma_{\mathcal{S}_\alpha}^2(\textbf{p}_{(X,Y)})
.
\end{eqnarray*}

\noindent This proves the claim \eqref{jRenan} and ends the proof of the Proposition  \ref{thejRen} . 

\end{proof}

\bigskip \noindent Note also that ,
 the relation \eqref{jte_salp} suggests that similar results of
Proposition \ref{corSalp} could be also extended to the JTE.\\

\bigskip \noindent For $\alpha>0,\ \ \alpha\neq 1$, denote   \begin{eqnarray*}
\label{tasz} A_{T,\alpha}(\textbf{p}_{(X,Y)})&=&%
\frac{1}{|1-\alpha|}A_{\alpha}(\textbf{p}_{(X,Y)})\\  
\text{and}\ \ \sigma_{T,\alpha}^2(\textbf{p}_{(X,Y)})&=&
\frac{1}{(1-\alpha)^2} \sigma_{\mathcal{S}_\alpha}^2(\textbf{p}_{(X,Y)})
%
.\end{eqnarray*}
 
\begin{proposition}
 \label{tsal}
 Under the same assumptions as in Proposition \ref{corSalp},  the following asymptotic results hold
\begin{eqnarray}\label{astsal}
&&\limsup_{n\rightarrow +\infty}\frac{\left\vert T_\alpha\left(\widehat{\textbf{p}}_{(X,Y)}^{(n)}\right)- T_\alpha\left(\textbf{p}_{(X,Y)}\right)\right\vert}{ a_{Z,n} }\leq A_{T,\alpha}(\textbf{p}_{(X,Y)}),\ \ \text{a.s.}\\
&&\sqrt{n}\left(T_\alpha\left(\widehat{\textbf{p}}_{(X,Y)}^{(n)}\right)- T_\alpha\left(\textbf{p}_{(X,Y)}\right)\right)\stackrel{\mathcal{D}}{ \rightsquigarrow} \mathcal{N}\left(0,\sigma_{T,\alpha}^2(\textbf{p}_{(X,Y)})\right),\ \ \text{as}\ \ n\rightarrow +\infty.\label{talis}
\end{eqnarray}
\end{proposition}
\begin{proof}The proof follows very simply from Proposition \ref{corSalp}, by writing  
\begin{eqnarray*}
T_\alpha( \widehat{\textbf{p}}_{(X,Y)}^{(n)})-T_\alpha\left(\textbf{p}_{(X,Y)}\right)&=&\frac{1}{1-\alpha}\left( \mathcal{S}_\alpha( \widehat{\textbf{p}}_{(X,Y)}^{(n)} )-\mathcal{S}_\alpha(\textbf{p}_{(X,Y)} ) \right).
\end{eqnarray*}
\end{proof}

\subsection{Asymptotic behavior of mutual information estimate}$\,$\\

\bigskip \noindent The following proposition establishes the almost sure convergence and the asymptotic normality of the estimator  $ I(\widehat{\textbf{p}}_{(X,Y)}^{(n)})$.

\begin{proposition}
\label{thmut}
 Under the same assumptions as in Proposition \ref{corSalp},  the following asymptotic results hold
 \begin{eqnarray}
 &&\label{asipz1}\limsup_{n\rightarrow +\infty}\frac{\left\vert I(\widehat{\textbf{p}}_{(X,Y)}^{(n)})- I(\textbf{p}_{(X,Y)})\right\vert}{ a_{Z,n} }\leq  A_H(\textbf{p}_{(X,Y)}),\ \ \text{a.s.}\\
&&\label{nipz}\sqrt{n}\left(I(\widehat{\textbf{p}}_{(X,Y)}^{(n)})- I(\textbf{p}_{(X,Y)}) \right)\stackrel{\mathcal{D}}{ \rightsquigarrow} \mathcal{N}(0,\sigma_H^2(\textbf{p}_{(X,Y)})),\ \ \text{as}\ \ n\rightarrow +\infty
 \end{eqnarray}where $A_H(\textbf{p}_{(X,Y)})$ and $\sigma_H^2(\textbf{p}_{(X,Y)})$ are given resp. by \eqref{asz} and \eqref{sigz}.
\end{proposition}

\begin{proof}

\noindent It is straightforward to write 
 \begin{eqnarray*}
&& I(\widehat{\textbf{p}}_{(X,Y)}^{(n)})-I(\textbf{p}_{(X,Y)})\\
&=&H(\textbf{p}_{(X,Y)})-H(\widehat{\textbf{p}}_{(X,Y)}^{(n)})-\sum_{(i,j)\in I\times J}\left(\widehat{p}_{Z,\delta_i^j}^{(n)}\log \left( p_{X,i}^{(n)}\, p_{Y,j}^{(n)}\right)-p_{Z,\delta_i^j}\log \left(p_{X,i}\, p_{Y,j}\right)\right)\\
 &=&H(\textbf{p}_{(X,Y)})-H(\widehat{\textbf{p}}_{(X,Y)}^{(n)})\\
 &&-\sum_{(i,j)\in I\times J}\widehat{p}_{Z,\delta_i^j}^{(n)}\left[\log \frac{ p_{X,i}^{(n)} }{p_{X,i}}+\log  \frac{ p_{Y,j}^{(n)}}{p_{Y,j}} \right]+ \left[\widehat{p}_{Z,\delta_i^j}^{(n)}-p_{Z,\delta_i^j} \right]\times 
 \log p_{X,i}\, p_{Y,j}.
 \end{eqnarray*} First, we have, for $n$ large enough and for any $(i,j)\in I\times J$  
 \begin{eqnarray*}
 && \widehat{p}_{Z,\delta_i^j}^{(n)}-p_{Z,\delta_i^j}\stackrel{a.s.}{\longrightarrow}0 \ \ \text{from}\ \ \eqref{pzn},\\
&& 1- \frac{ a_{X,n}}{p_{X,i}}\leq \frac{\widehat{p}_{X,i}^{(n)} }{p_{X,i}} \leq  1+ \frac{ a_{X,n}}{p_{X,i}},\ \ \text{and}\ \ 1- \frac{ a_{Y,n}}{p_{Y,j}}\leq \frac{\widehat{p}_{Y,j}^{(n)} }{p_{Y,j}} \leq  1+ \frac{ a_{Y,n}}{p_{Y,j}}.
 \end{eqnarray*}Hence, using that $\log(1+x)\approx x$, for $x$ small enough, we get, for any fixed $(i,j)\in  I\times J$ \begin{eqnarray*}
\left[\log \frac{ p_{X,i}^{(n)} }{p_{X,i}}+\log  \frac{ p_{Y,j}^{(n)}}{p_{Y,j}} \right] \approx   \frac{ a_{X,n}}{p_{X,i}}+ \frac{ a_{Y,n}}{p_{Y,j}}\stackrel{a.s.}{\longrightarrow} 0,\ \ \text{as}\ \ n\rightarrow+\infty,
 \end{eqnarray*}using \eqref{maxaxy}.\\

 \noindent 
Therefore, we have asymptotically
 \begin{eqnarray}\label{mi-jse}
I(\widehat{\textbf{p}}_{(X,Y)}^{(n)})-I(\textbf{p}_{(X,Y)})&\approx &H(\textbf{p}_{(X,Y)})-H(\widehat{\textbf{p}}_{(X,Y)}^{(n)}).
 \end{eqnarray} Finally, \eqref{asipz1} and \eqref{nipz} follow from the Proposition \ref{jent}.\\
 
 \noindent This ends the proof of the Proposition \ref{thmut}.
 
\end{proof}
\section{Statistic test of independence based on  mutual information}\label{testindep}

\noindent The proposed mutual information estimator is a natural test statistic for independence.
\noindent  Given two random variables $X$ and $Y$  with joint probability distribution 
 $\textbf{p}_{(X,Y)}=( p_{i,j})_{(i,j)\in [1,r]\times [1,s]}
 $, an hypothesis for testing the independence is $$H_0:\, I(\textbf{p}_{(X,Y)})= 0$$ versus $$H_1:\, I(\textbf{p}_{(X,Y)})>0.$$

\bigskip \noindent  From a random sample  $Z_1,Z_2,\cdots,Z_n$ according to $\textbf{p}_{(X,Y)}$, we compute the MI estimator $I(\widehat{ \textbf{p}}_{(X,Y)}^{(n)})$
.\\

\noindent Clearly, \eqref{asipz1} implies that under $H_0$, $I(\widehat{\textbf{p}}_{(X,Y)}^{(n)})\stackrel{a.s}{\longrightarrow} 0$, as  $n\rightarrow \infty$, and a classical result in statistics (see \cite{chri}, \cite{wilk},  and \cite{fan}
)
 establish that $2nI(\widehat{\textbf{p}}_{(X,Y)}^{(n)})$ approximately follows a $\chi^2$ distribution with
$(r-1)(s-1)$ degrees of freedom, for short
$$
2nI(\widehat{\textbf{p}}_{(X,Y)}^{(n)})\sim 
\chi_{(r-1)(s-1)}^2,$$for $n$ large.\\

\noindent \noindent Then, at significance level  $\alpha\in (0,1)$, 
 we reject the null hypothesis $H_0$
, when $2nI(\widehat{\textbf{p}}_{(X,Y)}^{(n)})$ is greater than the $(1-\alpha)$-th quantile of $\chi_{(r-1)(s-1)}^2$.

%

\section{Simulation}\label{simulat}
\noindent In this section, we start by providing a numerical example  to  illustrate 
 asymptotic  behavior of  the 
different  joint entropy  measures defined before. 
\\

 \noindent For simplicity consider two discretes random variables $X$ and $Y$ having each one two outcomes $x_1,x_2,x_3$ and $y_1,y_2$ and such that 

\begin{eqnarray*}
&&\mathbb{P}(X=x_1,Y=y_1)=	\frac{144}{205},\ \ \ \mathbb{P}(X=x_1,Y=y_2)=\frac{36}{205}\\
&&\mathbb{P}(X=x_2,Y=y_1)=\frac{16}{205},\ \  \mathbb{P}(X=x_2,Y=y_2)=\frac{9}{205}.
\end{eqnarray*}

\noindent So that the associated random variable $Z$, defined by \eqref{pijzk} and \eqref{zkpij}, is a discrete random variable whose probability distribution is that of a \textit{discrete Zipf distributions} $Z_{\beta,m}$ with parameter $\beta=2$ and $m=4
$. 
Its \textit{p.m.f.}  is defined by 
$$p_{Z,k}=\frac{k^{-\beta}}{\displaystyle \sum_{i=1}^{m}i^{-\beta}}\ \ \text{for}\ \ k =1,2,\cdots,m$$
where $\sum_{j=1}^mj^{-\beta}$ refers to the generalized harmonic function.\\

\bigskip \noindent We have 
\begin{eqnarray*}
&&H(\textbf{p}_X)= 0.3707947\,\ \text{nat},\ \  
H(\textbf{p}_Y)=0.5262899\,\ \text{nat}\\
&& H(\textbf{p}_{(X,Y)})
 = 0.8898576
\ \ \text{nat},\ \ \ R_{2}(\textbf{p}_{(X,Y)})= 0.6305886
 \ \ \text{nat}\ \\
 && T_{2}(\textbf{p}_{(X,Y)})= 0.4677216
 \ \text{nat},\ \ \text{and}\ \ \  I(\textbf{p}_{(X,Y)})= 0.0072269860\ \ \text{nat}.
\end{eqnarray*}
\noindent $Y$ is more uncertainty than $X$ and the pair $(X,Y)$ is less uncertainty than the discrete uniform distribution with range $[1,4]$ and which entropy is $\log 4= 1.386294$.  The variables 
$X$ and $Y$ seem not to have a lot of information in 
common,  only  $0.0072269860\ \ \text{nat}$ of information.\\
\vspace{2ex}
\begin{table}
\centering
\begin{tabular}{|c|c|c|c|c|c|}
\hline
&&&&\\
$(X,Y)$&$(x_1,y_1)$&$(x_1,y_2)$&$(x_2,y_1)$&$(x_2,y_2)$\\
&&&&\\
$p_{Z,k}$&$\frac{144}{205}$&$\frac{36}{205}$&$\frac{16}{205}$&$\frac{9}{205}$\\
&&&&\\
\hline
\end{tabular}
\vspace{3ex}
\caption{Joint \textit{p.m.f.} table of the random variable $Z$ with law $
\textbf{p}_{(X,Y)}$}
	\label{tab2}
\end{table} 
\noindent

\noindent  The \textsc{Table} \ref{tab2} defines the probability distribution $\textbf{p}_Z$, of $Z$.\\

\noindent In our applications we simulated i.i.d. samples of size $n$ ($n=100,200,\cdots,30000)$ according to $\textbf{p}_Z$, and computed the joint entropy estimates. 

%
%
\bigskip \noindent  \textsc{Figure} \ref{jsexyn},  concerns JSE estimate, 
 \textsc{Figure} \ref{jrexyn} concerns JRE 
 and JTE estimates (both of order $\alpha=2$)  
 , whereas \textsc{Figure} \ref{jmin} concerns  MI
estimate, all of the pair $(X,Y)$.\\

\noindent In each of these \textsc{F}igures, left panels 
  represent plot of the proposed entropy estimator, built from sample sizes of $n=100,200,\cdots,30000$, and the true entropy of the pair $(X,Y)$  (represented by horizontal black line). We observe 
 that when  
 the sample sizes $n$ increase, then the proposed estimator value converges almost surely to the true value. \\
 \noindent  Middle panels show the histogram of the sample and where the red line represents the plot of the theoretical normal distribution calculated
 from the same mean and the same standard deviation of the sample.\\
 \noindent Right panels concern the Q-Q plot of the sample which display the observed values against normally 
distributed data (represented by the red line). We observe that the  underlying 
distribution of the data is normal since the points fall along a straight line.\\


\begin{center}
\begin{figure}[H]
\includegraphics[scale=0.28]{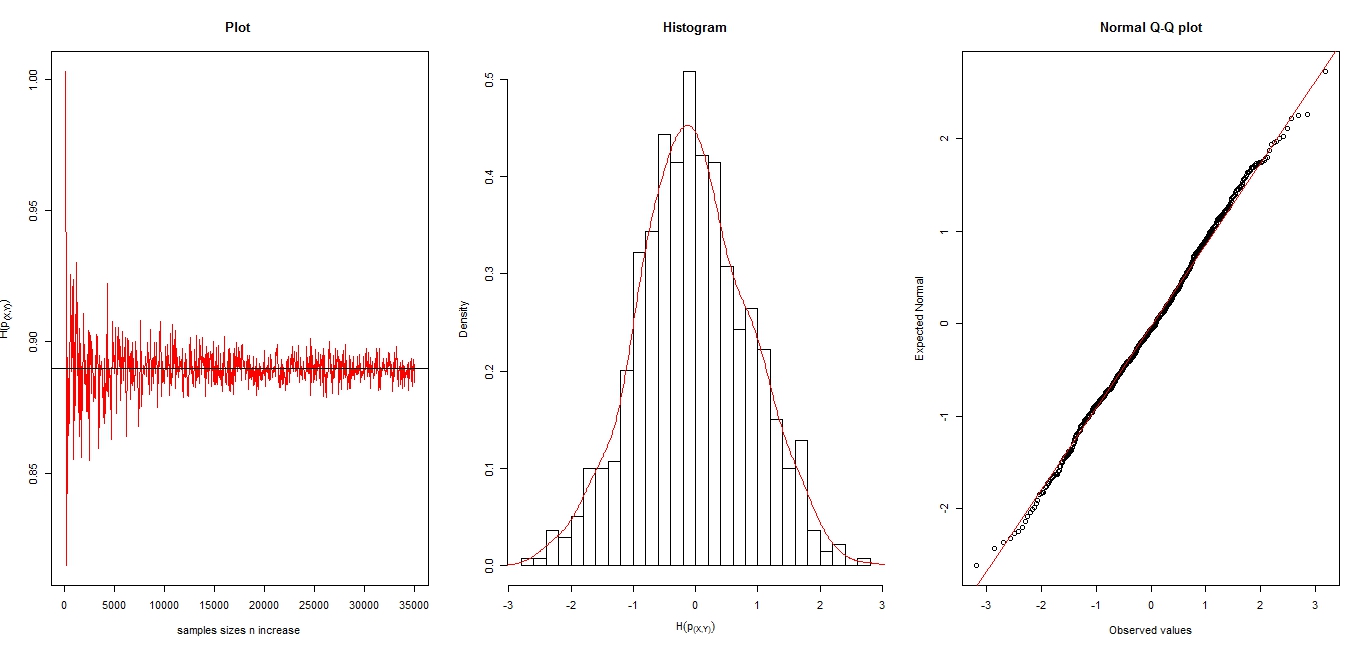} 
 \caption{Plot, histogram and normal Q-Q plot of JSE $ H(\widehat{\textbf{p}}_{(X,Y)}^{(n)})$.}\label{jsexyn}
\end{figure}
\begin{figure}[H]
\includegraphics[scale=0.26]{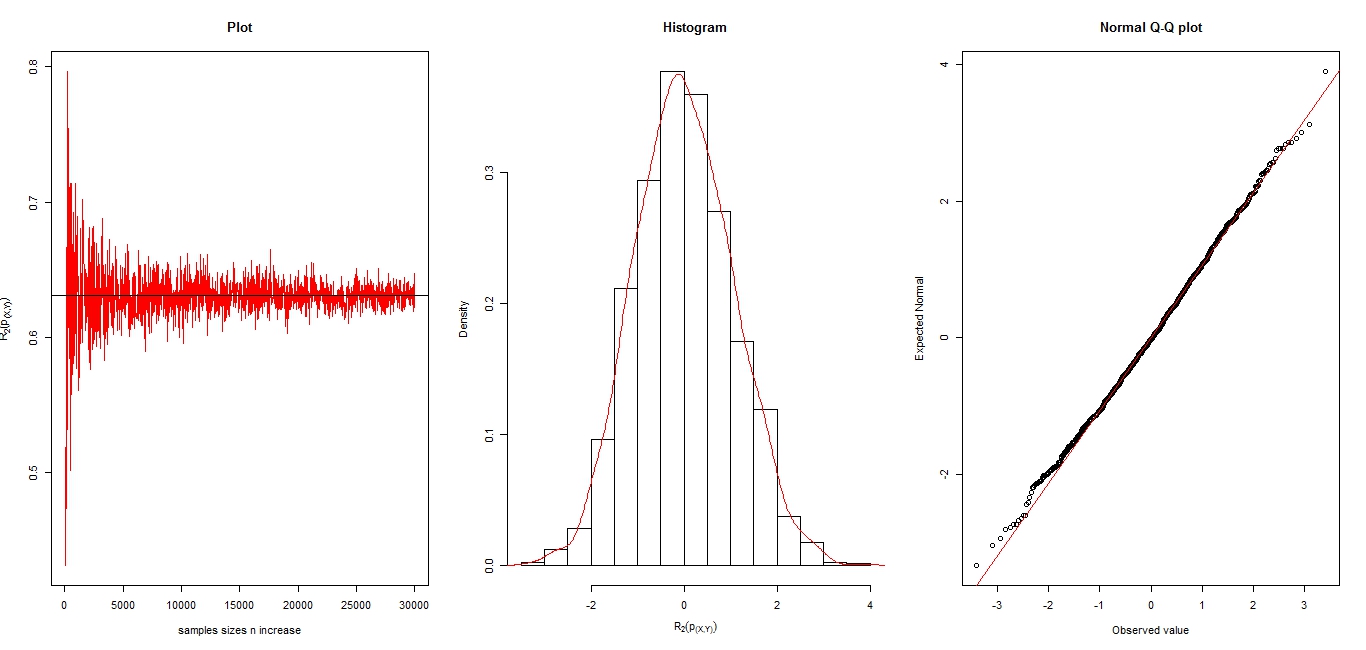} 
\includegraphics[scale=0.26]{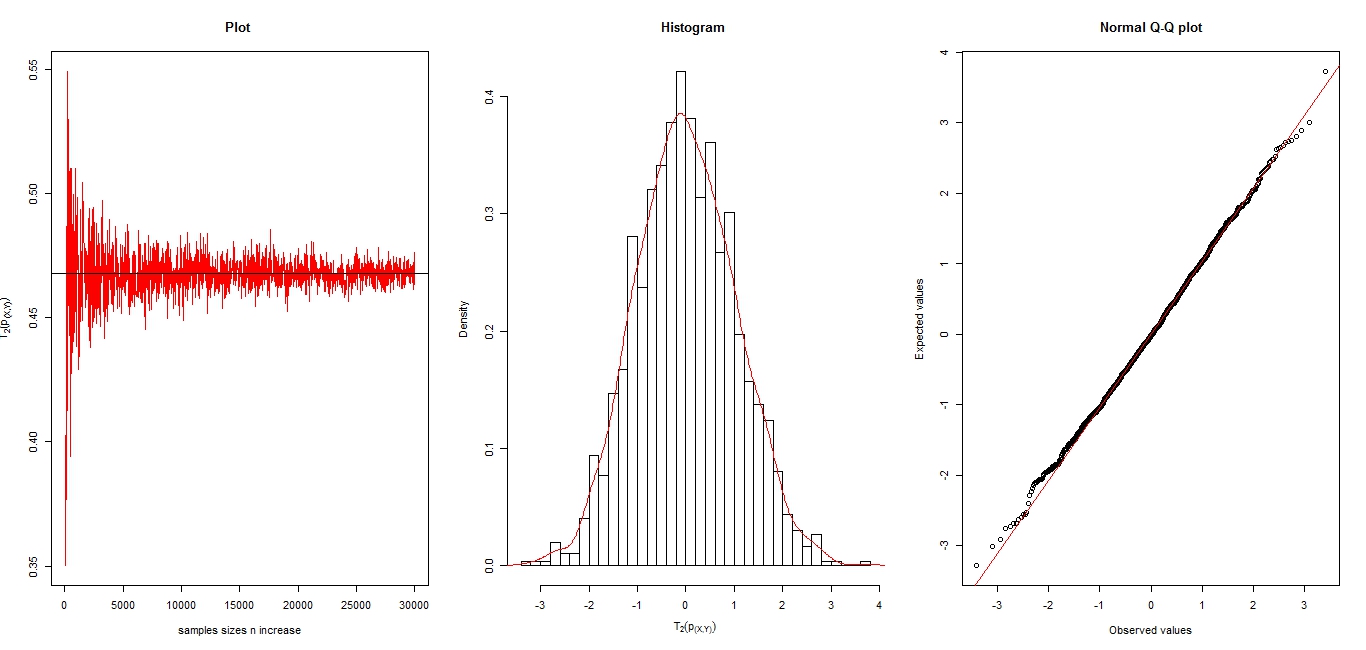}
 \caption{ 
Plots when samples sizes increase, histograms of $ R_{2}(\widehat{\textbf{p}}_{(X,Y)}^{(n)})$ and $ T_{2}(\widehat{\textbf{p}}_{(X,Y)}^{(n)})$. 
}\label{jrexyn}
\end{figure}

\begin{figure}[H]
\includegraphics[scale=0.26]{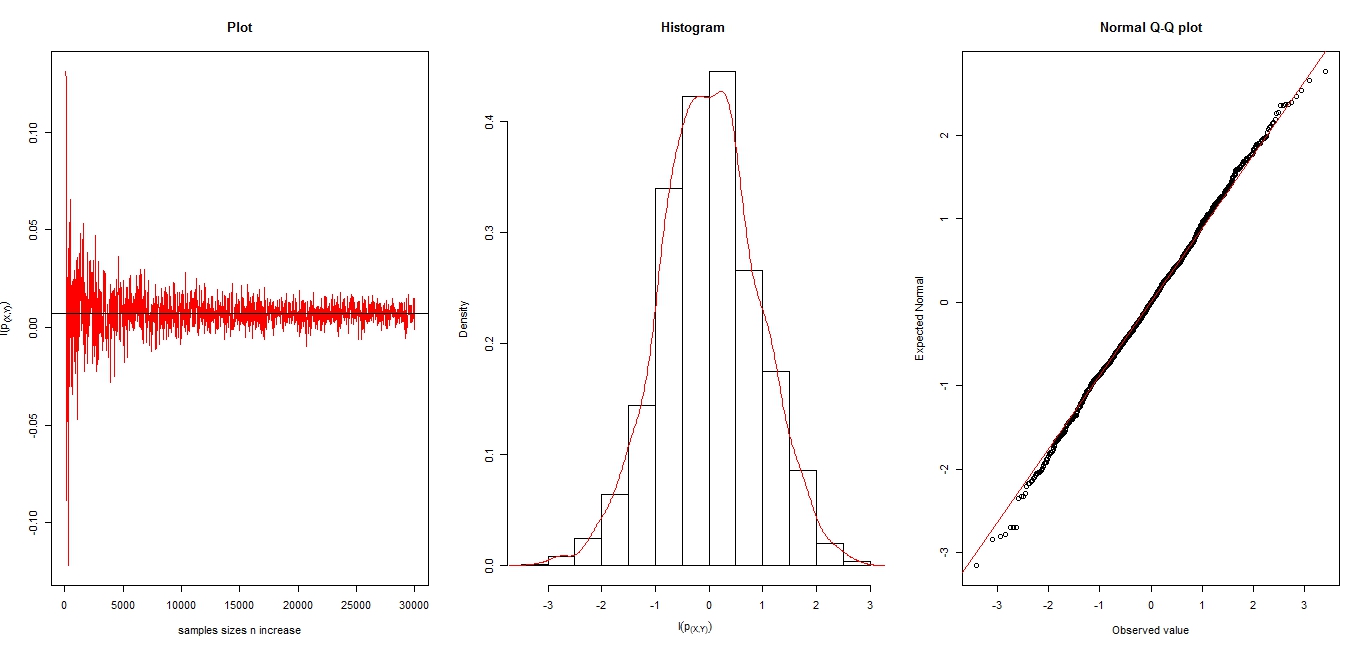} 
 \caption{Plot when samples sizes increase, histograms and normal Q-Q plot of MI estimate $ I(\widehat{\textbf{p}}_{(X,Y)}^{(n)})$.}\label{jmin} 
\end{figure}
\end{center}
%

\section{Conclusion}\label{conclus}
In this paper, we presented a new method for estimating the joint p.m.f. of a pair of discrete random variables. We adopted the plug-in method to construct estimates of joint shannon, Reyni and Tsallis entropies, and that of mutual information of a ordered pair of random variables. We established almost-sure rates of convergence and asymptotic normality of these 
estimators
.

\bibliographystyle{plain}

\begin{thebibliography}{10}
\bibitem[Carter(2014)]{carter}Carter, Tom (March 2014). \textit{An introduction to information theory and entropy} (PDF). Santa Fe.
\bibitem[R\'enyi(1960)]{ren2}
 R\'enyi, A. (1960), On measures of information and entropy, \textit{Proc. 4th Berkeley
Symposium on Mathematics, Statistics and Probability}, pp 547-561.
\bibitem[ Kraskov et al.(2004)]{kras} Kraskov A, St\'ogbauer H, Grassberger P (2004). \textit{Estimating mutual information}. Phys Rev E 69:066138.

\bibitem[Philippatos $\&$ Wilson(1972)]{phil}
Philippatos, G.C.; Wilson, C.J. (1972). Entropy, market risk, and the selection of efficient portfolios. \textit{Appl. Econ.}, \textbf{ 4}, pp. 209–220.
\bibitem[Moon and al.(2017)]{moon2} Moon KR, Sricharan K, Hero AO (2017). Ensemble estimation of mutual information. \textit{IEEE International Symposium on Information Theory (ISIT)}, eds Durisi G,
Studer C (IEEE, Aachen, Germany), pp 3030–3034.
\bibitem[Timme $\&$ Lapish (2018)]{timm}
Timme NM, Lapish C.(2018). A Tutorial for Information Theory in Neuroscience. \textit{eNeuro.} \textbf{5}(3)

\bibitem[Krishnaswamy et al.(2014)]{krishn} Krishnaswamy, Matthew H Spitzer, Michael Mingueneau, Sean C Bendall, Oren Litvin, Erica
Stone, Dana Pe\'er, and Garry P Nolan (2014). Conditional density-based analysis of t cell signaling in single-cell
data. \textit{Science}, 346(6213):1250689.
\bibitem[Liu et al.(2012)]{liu}
 H. Liu, L. Wasserman, and J. D. Lafferty(2012), Exponential concentration for mutual information estimation with application to forests, \textit{in Advances in Neural Information Processing Systems}, pp. 2537-2545.
\bibitem[Lewi et al.(2006)]{lewi}Lewi, R. Butera, and L. Paninski (2006). Real-time adaptive information-theoretic optimization of neurophysiology experiments, \textit{in Advances in
Neural Information Processing Systems}, pp. 857-864.
\bibitem[P\'al et al.(2010)]{pal} D. P\'al, B. P\'oczos, and C. Szepesv\'ari (2010). Estimation of R\'enyi entropy and
mutual information based on generalized nearest-neighbor graphs, in
\textit{Advances in Neural Information Processing Systems}, pp. 1849-1857.
\bibitem[Christensen (1997)]{chri}
R. Christensen (1997). \textit{Log-linear Models and Logistic Regression}. Springer,
New York.
\bibitem[Reshef et al.(2011)]{dave}  David N Reshef, Yakir A Reshef, Hilary K Finucane, Sharon R Grossman, Gilean McVean, Peter J
Turnbaugh, Eric S Lander, Michael Mitzenmacher, and Pardis C Sabeti (2011). Detecting novel associations in
large data sets.\textit{ science}, 334(6062), pp. 1518–1524.


\bibitem[Rieke(1999)]{fred}
Fred Rieke. Spikes: exploring the neural code. MIT press, 1999.

\bibitem[Schneidman et al.(2003)]{schne} E. Schneidman, W. Bialek, and M. J. B. II (2003). An information theoretic
approach to the functional classification of neurons, \textit{Advances in Neural
Information Processing Systems}, vol.\textbf{15}, pp. 197-204.

\bibitem[Walters et al.(2009)]{walt} Janett Walters-Williams and Yan Li. (2009). Estimation of mutual information:
A survey. \textit{In International Conference on Rough Sets and Knowledge Technology}
(RSKT'08). 389–396.
\bibitem[Khan et al.(2007)]{khan} Shiraj K., Sharba B., Auroop R. G., Sunil S., David J.
E., Vladimir P., and George O. (2007). Relative
performance of mutual information estimation methods for quantifying the
dependence among short and noisy data. \textit{Phys. Rev. E} 76, 2 (2007), 026209.
\bibitem[Sricharan et al.(2013)]{sric}
K. Sricharan, D. Wei, and A. O. Hero (2013). Ensemble estimators for multivariate entropy estimation.
\textit{Information Theory, IEEE Transactions on}, 59(7): 4374–4388.


\bibitem[Antos and Kontoyiannis(2001)]{antos}
Antos A. and Kontoyiannis I.(2001). Convergence Properties of Functional Estimates for Discrete Distributions.
\textit{Random Structures and Algorithms}, 19(3‐4), 163-193, October 2001.
https://doi.org/10.1002/rsa.10019
\bibitem[Deemat (2013)]{deem}Deemat C Mathew (2013). Nonparametric Estimation of Mutual Information 
and Test for Independence 
 \textit{International Journal of Statistika and Mathematika}, ISSN : 2277- 2790 E-ISSN: 2249-8605, Volume 5, Issue 2, 2013 pp 27-30.
 
\bibitem[Gao et al.(2017a)]{gaow}
 Gao W.,  Oh S., and  Viswanath P.(2017). Demystifying fixed $k$-nearest neighbor
information estimators. \textit{In Information Theory (ISIT)}, 2017 IEEE International Symposium on,
pages 1267–1271. IEEE.

\bibitem[Gao et al.(2017)]{gao}Gao W, Kannan S, Oh S, Viswanath P (2017) Estimating mutual information for discrete-continuous mixtures. \textit{Advances in Neural Information Processing Systems}, eds Guyon I, et al. (Curran Associates, Inc., Red Hook, NY), Vol 30, pp 5986-5997.


\bibitem[Goebel et al.(2005)]{goeb}
B. Goebel, Z. Dawy, J. Hagenauer, and J.C. Mueller (2005). An approximation to the distribution of finite sample size mutual information estimates. \textit{IEEE International Conference on Communications, 2005. ICC}. DOI: 10.1109/ICC.2005.1494518.
\bibitem[Xianli et al.(2018)]{xian}
Xianli Zeng, Yingcun Xia, and Howell Tong. Jackknife approach to the estimation
of mutual information.
\textit{
PNAS October 2, 2018 115 (40) 9956-9961}; first published September 17, 2018 https://doi.org/10.1073/pnas.1715593115



\bibitem[Beknazaryan et al.(2019)]{bekn}
Aleksandr Beknazaryan, Xin Dang, Hailin Sang
(2019).On mutual information estimation for mixed-pair random variables. \textit{Statistics and  Probability Letters} 148 (2019) 9-16nhttps://doi.org/10.1016/j.spl.2018.12.011.
\bibitem[Hall(1987)]{hall} Hall,P. (1987). On Kullback-Leibler loss and density estimation. \textit{The Annals of Statistics}, Vol.15(4), pp.1491-1519.

 \bibitem[Singh and Poczos (2014)]{singh} Singh S. and Poczos, B. (2014). Generalized Exponential Concentration Inequality for R\'{e}nyi Divergence Estimation. \textit{Journal of Machine Learning Research}.Vol.6. Carnegie Mellon University. 
 \bibitem[Krishnamurthy \textit{et al.}(2014)]{kris} Akshay K., Kirthevasan K., Poczos B., and Wasserman, L.(2014). Nonparametric Estimation of R\'{e}nyi Divergence and Friends. \textit{Journal of Machine Learning Research} Workshop and conference Proceedings, 32. Vol.3, pp. 2. 
 \bibitem[Ba et al.(2019)]{ba5}
 Ba, A.D, Lo G.S.(2019), Divergence Measures Estimation and Its Asymptotic Normality Theory in the discrete case. \textit{European Journal of Pure and Applied Mathematics}, Vol.12, No.3, 790-820.
\bibitem[Lo (2016)]{ips-wcia-ang}
Lo, G.S.(2016). Weak Convergence (IA). Sequences of random vectors.
SPAS Books Series. Saint-Louis, Senegal - Calgary, Canada. Doi :
10.16929/sbs/2016.0001. Arxiv : 1610.05415. ISBN : 978-2-9559183- 1-9.

\bibitem[Wilks(1938)]{wilk}
Wilks SS(1938). The large-sample distribution of the likelihood ratio for testing composite
hypotheses. \textit{Ann Math Stat}, 9, 60-2
\bibitem[Fan et al.(2000)]{fan}
Fan J, Hung HN, Wong WH(2000). Geometric  understanding of likelihood ratio statistics. \textit{J Am Stat Assoc}. 95, pp. 836-41.



%
%
%
%
%
%
%
%
%
%



 




































































\end{thebibliography}

\end{document}